\documentstyle[12pt]{article}
\input math_macros.tex

\def\ref#1{$^{#1)}$}
\begin{document}
\begin{titlepage}
\begin{center}
June 22, 1992     \hfill    LBL-32491 (Revised)\\

\vskip .05in

{\large \bf Magnetic Monopoles As a New Solution to Strong CP Problem}
\footnote{This work was supported by the Director, Office of Energy
Research, Office of High Energy and Nuclear Physics, Division of High
Energy Physics of the U.S. Department of Energy under Contract
DE-AC03-76SF00098.}
\vskip .05in
Huazhong Zhang\\[.1in]
{\em Theoretical Physics Group\\
     Lawrence Berkeley Laboratory\\
     MS 50A-3115, 1 Cyclotron Road\\
     Berkeley, California 94720}
\end{center}

\begin{abstract}
A non-perturbative solution to strong CP problem is proposed. It is shown that
the gauge orbit space with gauge potentials and gauge tranformations restricted
on the space boundary in non-abelian gauge theories with a $\theta$ term has a
magnetic monopole structure if there is a magnetic monopole in the ordinary
space. The Dirac's quantization condition in the corresponding quantum theories
ensures that the vacuum angle $\theta$ in the gauge theories must be quantized.
The quantization rule is derived as $\theta=2\pi/n~(n\neq 0)$ with n being the
topological charge of the magnetic monopole. Therefore, we conclude that the
strong CP problem is automatically solved non-perturbatively with the existence
of a magnetic monopole of charge $\pm 1$ with $\theta=\pm 2\pi$. This is also
true when the total magnetic charge of monopoles are very large
($|n|\geq 10^92\pi$) if it is consistent with the abundance of magnetic
monopoles. This implies that the fact that the strong CP violation can be only
so small or vanishing may be a signal for the existence of magnetic monopoles.
\end{abstract}
\end{titlepage}
\newpage
\renewcommand{\thepage}{\arabic{page}}
\setcounter{page}{1}
Since the discovery of Yang-Mills theories$^1$, the non-perturbative effects of
gauge theories have played one of the most important roles in particle physics.
It is known that, in non-abelian gauge theories a Pontryagin or $\theta$ term,
\begin{equation}
{\cal L}_{\theta}=\frac{\theta}{32{\pi}^2}\epsilon^{\mu\nu\lambda\sigma}F^
a_{\mu\nu}F^a_{\lambda\sigma},
\end{equation}
can be added to the Lagrangian density of the system due to instanton$^2$
effects in gauge theories. The $\theta$ term can induce CP violations. An
interesting fact is that the $\theta$ angle in QCD can be only very small
(${\theta\leq}10^{-9}$) or vanishing$^3$. Where in our discussions of QCD,
$\theta$ is used to denote $\theta+arg(detM)$ effectively with M being the
quark mass matrix, when the effects of electroweak interactions are included.
One of the most interesting understanding of the strong CP problem has been
the assumption of an additional Peccei-Quinn $U(1)_{PQ}$ symmetry$^4$, but the
observation has not given$^3$ evidence for the axions$^5$ needed in this
approach. Thus the other possible solutions to this problem are of fundamental
interest.

In this paper, we will extend the method of Wu and Zee$^6$ for the discussions
of the effects of the Pontryagin term in pure Yang-Mills theories in the gauge
orbit spaces in the Schrodinger formulation. This formalism has also been used
with different methods to derive the mass parameter quantization in
three-dimensional Yang-Mills theory with Chern-Simons term$^{6-7}$.  Wu and Zee
showed$^6$ that the Pontryagin term induces an abelian background field or an
abelian structure in the gauge configuration space of the Yang-Mills theory.
In our discussions, we will consider the case with the existence of a magnetic
monopole. We will show that magnetic monopoles$^{8-9}$ in space will induce an
abelian gauge field with non-vanishing field strength in gauge configuration
space, and magnetic flux through a two-dimensional sphere in the induced gauge
orbit space is non-vanishing. Then, Dirac condition$^{8-9}$ in the
corresponding quantum theories leads to the result that the relevant vacuum
angle $\theta$ must be quantized as $\theta=2\pi/n$ with n being the
topological charge of the monopole to be generally defined. Therefore, the
strong CP problem can be solved with the existence of magnetic monopoles.
To the knowledge of the author, such an interesting result has never been given
before in the literature.

We will now consider the Yang-Mills theory with the existence of a magnetic
monopole at the origin. As we will see that an interesting feature in our
derivation is that we will use the Dirac quantization condition both in the
ordinary space and restricted gauge orbit space to be defined. The Lagrangian
of the system is given by
\begin{equation}
{\cal L}={\int}d^4x\{-\frac{1}{4}F^a_{\mu\nu}F^{a\mu\nu}+\frac{\theta}{32\pi^2}
\epsilon^{\mu\nu\lambda\sigma}F^a_{\mu\nu}F^{a\lambda\sigma}\}.
\end{equation}
We will use the Schrodinger formulation and the Weyl gauge $A^0=0$. The
conjugate
momentum corresponding to $A^a_i$ is given by
\begin{equation}
\pi^a_i=\frac{\delta{\cal L}}{\delta\dot{A}^a_i}=\dot{A}^a_i+\frac{\theta}
{8\pi^2}\epsilon_{ijk}F^a_{jk}.
\end{equation}
In the Schrodinger formulation, the system is similar to the quantum system
of a particle with the coordinate $q_i$ moving in a gauge field $A_i(q)$ with
the correspondence$^{6-7}$
\begin{eqnarray}
q_i(t)\rightarrow A^a_i({\bf x},t),\\
A_i(q)\rightarrow {\cal A}^a_i({\bf A}({\bf x})),
\end{eqnarray}
where
\begin{eqnarray}
{\cal A}^a_i({\bf A}({\bf x}))=\frac{\theta}{8\pi^2}\epsilon_{ijk}F^a_{jk}.
\end{eqnarray}
Thus there is a gauge structure with gauge potential ${\cal A}$ in this
formalism within a gauge theory with the ${\theta}$ term included. Note that
in our discussion with the presence of a magnetic monopole, the gauge
potential ${\bf A}$ outside the monopole generally need to be understood as
well defined in each local coordinate region. In the overlapping regions, the
separate gauge potentials can only differ by a well-defined gauge
transformation$^9$. In fact, single-valuedness of the gauge function
corresponds to the Dirac quantization condition$^9$. For a given r, we can
choose two extended semi-spheres around the monopole, with $\theta
\in[\pi/2-\delta,\pi/2+\delta] (0<\delta<\pi/2)$ in the overlapping region,
where the $\theta$ denotes the $\theta$ angle in the spherical polar
coordinates. For convenience, we will use differential forms$^{10}$ in our
discussions, where $A=A_idx^i, F=\frac{1}{2}F_{jk}dx^jdx^k,$ with $F=dA+A^2$
locally. For our purpose to discuss about the effects of the abelian gauge
structure on the quantization of the vacuum angle, we will now briefly clarify
the relevant topological results needed, then we will realize the topological
results explicitly.

With the constraint of Gauss' law, the quantum theory in this formalism is
described in the gauge orbit space ${\cal U}$/${\cal G}$ which is a quotient
space of the gauge configuration space ${\cal U}$ with gauge potentials
connected by gauge transformations in each local coordinate region regarded as
equivalent. Where ${\cal G}$ denotes the space of continuous gauge
transformations, and each gauge potential as an element in ${\cal U}$ may be
defined up to a gauge transformation in the overlapping regions. Now consider
the following exact homotopy sequence$^{11}$:
\begin{equation}
\Pi_N({\cal U})\stackrel{P_*}{\longrightarrow}\Pi_N({\cal U}/{\cal G})
\stackrel{\Delta_*}{\longrightarrow}\Pi_{N-1}({\cal G})\stackrel{i_*}
{\longrightarrow}\Pi_{N-1}({\cal U}) ~ (N\geq 1).
\end{equation}
Note that homotopy theory has also been used to study the global
gauge anomalies $^{12-19}$, especially by using
extensively the exact homotopy sequences and in terms of James numbers of
Stiefel manifolds$14-19$. One can easily see that
${\cal U}$ is topologically trivial,
thus $\Pi_N({\cal U})=0$ for any N. Since
the interpolation between any two gauge potentials $A_1$ and $A_2$
\begin{equation}
A_t=tA_1+(1-t)A_2
\end{equation}
for any real t is in ${\cal U}$ (Theorem 7 in Ref.9, and Ref.6). since $A_t$ is
transformed as a gauge potential in each local coordinate region, and in an
overlapping region, both $A_1$ and $A_2$ are gauge potentials may be defined up
to a gauge transformation, then $A_t$ is a gauge potential which may be defined
up to a gauge transformation, namely, $A_t\in{\cal U}$. Thus, we have
\begin{equation}
0\stackrel{P_*}{\longrightarrow}\Pi_N({\cal U}/{\cal G})
\stackrel{\Delta_*}{\longrightarrow}\Pi_{N-1}({\cal G})\stackrel{i_*}
{\longrightarrow}0 ~ (N\geq 1).
\end{equation}
This implies that
\begin{equation}
\Pi_N({\cal U}/{\cal G})\cong\Pi_{N-1}({\cal G})~ (N\geq 1).
\end{equation}

As we will show that in the presence of a magnetic monopole, the topological
properties of the system are drastically different. This will give important
consequences in the quantum theory.
Actually, it is interesting to note more generally that the topological results
in Eq.(9-10) are true if $\cal U$ and $\cal G$ are the corresponding induced
spaces with A and g restricted to certain region of the ordinary space,
especially the 2-sphere $S^2$ as the space boundary since the restricted gauge
configuration space $\cal U$ is topologically trivial. This is in fact
the the relevant case in our discussion, since only the integrals on the space
boundary $S^2$ are relevant in the quantization equation as we will see. We
will call the induced spaces of $\cal U$, $\cal G$ and ${\cal U}/{\cal G}$ when
A and g are restricted on the space boundary $S^2$ as restricted gauge
configuration space, restricted gauge orbit space and restricted gauge
transformation space respectively, and restricted spaces collectively. Now for
the restricted spaces, the main topological result we will use is given by
\begin{equation}
\Pi_2({\cal U}/{\cal G})\cong\Pi_1({\cal G}),
\end{equation}
The condition $\Pi_2({\cal U}/{\cal G})\neq 0$ corresponds to the existence of
a magnetic monopole in the restricted gauge
orbit space. In the usual unrestricted case based on the whole compactified
space M as that for pure Yang-Mills theory, there can not be monopole structure
constructed. We will first show that in this case ${\cal F}\neq 0$, and then
demonstrate explicitly that the magnetic flux $\int_{S^2}\hat{\cal F}\neq 0$
can be nonvanishing in the restricted gauge orbit space, where $\hat{\cal F}$
denotes the projection of $\cal F$ into the restricted gauge orbit space.

Denote the differentiation with respect to
space variable ${\bf x}$ by d, and the differentiation with respect to
parameters $\{t_i\mid i=1,2...\}$ which {\bf A}({\bf x}) may depend on in the
gauge configuration space by $\delta$, and assume $d\delta+\delta d$=0. Then,
similar to $A=A_{\mu}dx^{\mu}$ with $\mu$ replaced by a, i, ${\bf x}$,
$A=A^a_iL^adx^i, F=\frac{1}{2}F^a_{jk}L^adx^jdx^k$ and
$tr(L^aL^b)=-\frac{1}{2}{\delta}^{ab}$ for a basis
$\{L^a\mid a=1, 2,...,rank(G)\}$ of the Lie algebra of the gauge group G,
the gauge potential in the gauge configuration space is given by
\begin{equation}
{\cal A}=\int d^3x{\cal A}^a_i({\bf A}({\bf x}))\delta A^a_i(\bf x).
\end{equation}
Using Eq.(6), this gives
\begin{equation}
{\cal A}=\frac{\theta}{8\pi^2}\int d^3x\epsilon_{ijk}F^a_{jk}({\bf x})\delta
A^a_i({\bf x})=-\frac{\theta}{2\pi^2}\int_M tr(\delta AF),
\end{equation}
with M being the space manifold. With $\delta F=-D_A(\delta A)=
-\{d(\delta A)+A\delta A-\delta AA\}$, we have topologically
\begin{equation}
{\cal F}=\delta{\cal A}=\frac {\theta}{2\pi^2}\int_Mtr[\delta AD_A(\delta A)]
=\frac {\theta}{4\pi^2}\int_Mdtr(\delta A\delta A)
=\frac {\theta}{4\pi^2}\int_{\partial M}tr(\delta A\delta A).
\end{equation}
Usually, one may assume $A\rightarrow 0$ faster than 1/r as $\bf x\rightarrow
0$
, then$^6$ this would give ${\cal F}=0$. However, this is not the case in the
presence of a magnetic monopole. Asymptotically, a monopole may generally give
a field strength of the form$^{8-9,20}$
\begin{equation}
F_{ij}=\frac{1}{4\pi r^2}\epsilon_{ijk}({\bf {\hat r}})_kG({{\bf \hat r}}),
\end{equation}
with $\bf {\hat r}$ being the unit vector for {\bf r}, and this gives
$A\rightarrow O(1/r)$ as $\bf x\rightarrow 0$. Thus, one can see
easily that a magnetic monopole can give a nonvanishing field strength $\cal F$
in the gauge configuration space. To evaluate the $\cal F$, one needs to
specify the space boundary $\partial M$ in the presence of a magnetic monopole.
we now consider the case that the magnetic monopole does not generate a
singularity in the space. In fact, this is so when monopoles appear as a smooth
solution of a spontaneously broken gauge theory similar to 't Hooft Polyakov
monopole$^8$. For example, it is known that$^{21}$ there are monopole solutions
in the minimal SU(5) model. Then, the space boundary may be regarded as a large
2-sphere $S^2$ at spatial infinity. For our purpose, we actually only need to
evaluate the projection of $\cal F$ into the gauge orbit space.

In the gauge orbit space, a gauge potential can be written in the form of
\begin{equation}
A=g^{-1}ag+g^{-1}dg,
\end{equation}
for an element a $\in{\cal U}/{\cal G}$ and a gauge function $g\in{\cal G}$.
Then the projection of a form into the gauge orbit space contains only terms
proportional to $(\delta a)^n$ for integers n. We can now write
\begin{equation}
\delta A=g^{-1}[\delta a-D_a(\delta gg^{-1})]g.
\end{equation}
Then we obtain
\begin{equation}
{\cal A}=-\frac{\theta}{2\pi^2}\int_M tr(f\delta a)
+\frac{\theta}{2\pi^2}\int_M tr[fD_a(\delta gg^{-1})],
\end{equation}
where $f=da+a^2$. With some calculations, this can be simplified as
\begin{equation}
{\cal A}=\hat{\cal A}
+\frac{\theta}{2\pi^2}\int_{S^2}tr[f\delta gg^{-1}],
\end{equation}
where
\begin{equation}
\hat{\cal A}=-\frac{\theta}{2\pi^2}\int_M tr(f\delta a),
\end{equation}
is the projection of $\cal A$ into the gauge orbit space. Similarly, we have
\begin{equation}
{\cal F}=\frac {\theta}{4\pi^2}\int_{S^2}
tr\{[\delta a-D_a(\delta gg^{-1})][\delta a-D_a(\delta gg^{-1})]\}
\end{equation}
or
\begin{equation}
{\cal F}=\hat{\cal F}-\frac {\theta}{4\pi^2}\int_{S^2}
tr\{\delta aD_a(\delta gg^{-1})+D_a(\delta gg^{-1})\delta a
-D_a(\delta gg^{-1})D_a(\delta gg^{-1})\},
\end{equation}
where
\begin{equation}
\hat{\cal F}=\frac {\theta}{4\pi^2}\int_{S^2}tr(\delta a\delta a).
\end{equation}
Now all our discussions will be based on the restricted spaces.
To see that the flux of $\hat{\cal F}$ through a closed surface in the
restricted gauge orbit space ${\cal U}/{\cal G}$ can be nonzero, we will
construct a 2-sphere in it. Consider an element $a\in{\cal U}/{\cal G}$, and a
loop in $\cal G$. The set of all the gauge potentials obtained by all the gauge
transformations on $a$ with gauge functions on the loop then forms a loop
$C^1$ in the gauge configurations space $\cal U$. Obviously, the $a$ is the
projection of the loop $C^1$ into ${\cal U}/{\cal G}$.
Now since $\Pi_1({\cal U})=0$ is trivial, the loop $C^1$ can be continuously
extented to a two-dimensional disc $D^2$ in the $\cal U$ with
$\partial D^2=C^1$, then obviously, the projection of the $D^2$ into the
gauge orbit space is topologically a 2-sphere $S^2\subset{\cal U}/{\cal G}$.
With the Stokes' theorem in the gauge configuration space, We now have
\begin{equation}
\int_{D^2}{\cal F}=\int_{D^2}\delta{\cal A}=\int_{C^1}{\cal A}.
\end{equation}
Using Eqs.(19) and (24) with $\delta a=0$ on $C^1$, this gives
\begin{equation}
\int_{C^1}{\cal A}
=\frac{\theta}{2\pi^2}tr\int_{S^2}\int_{C^1}[f\delta gg^{-1}].
\end{equation}
Thus, the projection of the Eq.(26) to the gauge orbit space gives
\begin{equation}
\int_{S^2}\hat{\cal F}
=\frac{\theta}{2\pi^2}tr\int_{S^2}\{f\int_{C^1}\delta gg^{-1}\},
\end{equation}
where note that in the two $S^2$ are in the gauge orbit space and the ordinary
space respectively. We have also obtained this by verifying that
\begin{equation}
\int_{D^2}tr\int_{S^2}
tr\{\delta aD_a(\delta gg^{-1})+D_a(\delta gg^{-1})\delta a
-D_a(\delta gg^{-1})D_a(\delta gg^{-1})\}=0,
\end{equation}
or the projection of $\int_{D^2}{\cal F}$ gives $\int_{S^2}\hat{\cal F}$.

In quantum theory, Eq.(26) corresponds to the topological result
$\Pi_2({\cal U}/{\cal G})\cong\Pi_{1}({\cal G})$ on the restricted spaces.
The discussion about the Hamiltonian equation in the schrodinger formulation
will be similar to that in Refs.6 and 7 including the discussions for the
three-dimensional Yang-Mills theories with a Chern-Simons term. We only need
the Dirac quantization condition here for our purpose. In the gauge orbit
space,
the Dirac quantization condition gives
\begin{equation}
\int_{S^2}\hat{\cal F}=2\pi k,
\end{equation}
with k being integers. Now let $f$ be the field strength 2-form for the
magnetic monopole. The quantization condition is now given by$^{20}$
\begin{equation}
exp\{\int_{S^2}f\}=exp\{G_0\}=exp\{4\pi\sum_{i=1}^{r}\beta^{i}H_{i}\}\in Z.
\end{equation}
Where $G_0$ is the magnetic charge up to a conjugate transformation by a group
element, $H_i$ (i=1, 2,...,r=rank(G)) form a basis for the Cartan subalgebra of
the gauge group with simple roots $\alpha_i$ (i=1,2,...,r). We need non-zero
topological value to obtain quantization condition for $\theta$. One way to
obtain non-zero value for Eq.(26) is to consider g(t) in the following U(1)
subgroup on the $C^1$
\begin{equation}
g(t)=exp\{4\pi mt\sum_{i,j}\frac{(\alpha_i)^jH_j}{<\alpha_i,\alpha_i>}\},
\end{equation}
with m being integers and $t\in [0,1]$. In fact, m should be identical to k
according to our topological result
$\Pi_2({\cal U}/{\cal G})\cong\Pi_{1}({\cal G})$. The k and m are the
topological numbers on each side. Thus, we obtain
\begin{equation}
\theta=\frac{2\pi}{n}~(n\neq0).
\end{equation}
Where we define generally the topological charge of the magnetic monopole as
\begin{equation}
n=-2<\delta,\beta>=-2\sum_i<\lambda_i,\beta>,
\end{equation}
which must be an integer$^{20}$. Where
\begin{equation}
\delta=\sum_i\frac{2\alpha_i}{<\alpha_i,\alpha_i>}=\sum_i\lambda_i,
\end{equation}
with the $\lambda_i$ being the fundamental weights of the Lie algebra, the
minus sign is due to our normalization convention for Lie algebra generators.

Therefore, we conclude that in the presence of magnetic monopoles with
topological charge $\pm 1$, the vacuum angle of non-abelian gauge theories
must be $\pm 2\pi$, the existence of such magnetic monopoles gives a solution
to the strong CP problem. But CP cannot be exactly conserved in this case since
$\theta=\pm 2\pi$ correspond to two different physical sysytems. The existence
of many monopoles can ensure $\theta\rightarrow 0$, and the strong CP problem
may also be solved. In this possible solution to the strong CP problem with
$\theta\leq 10^{-9}$, the total magnetic charges present are
$|n|\geq 2\pi 10^{9}$. This may possiblely be within the abundance allowed by
the ratio of monopoles to the entropy$^{22}$, but with the possible existence
of both monopoles and anti-monopoles, the total number of magnetic monopoles
may be larger than the total magnetic charges. Generally, one needs to ensure
that the total number is consistent with the experimental results on the
abundance of monopoles. The $n=\pm 2$ may also possibilely solve CP if it is
consistent with the experimental observation.

In the above discussions, we consider the case that magnetic monopole generates
no singularity in the space. If we consider the magnetic monopole as a
singularity, then with the space outside the monopole the two opposite boundary
contributions are cancelled in the relevant integrations since each inward
small sphere arround the monopole for removing the singularity effectively
gives a contribution of the opposite topological charge. Therefore, only
non-singular magnetic monopoles may provide the solution to the strong CP
problem.

Moreover, note that our conclusions are also true if we add an additional
$\theta$ term in QED with the $\theta$ angle the same as the effective
$\theta$ in QCD if there exist Dirac monopoles as color singlets, or a
non-abelian monopoles with magnetic charges both in the color SU(3) and
electromagnetic U(1). Then a explanation of such a QED $\theta$ term is needed.
The effect of a term proportional to $\epsilon^{\mu\nu\lambda\sigma}F_{\mu\nu}
F_{\lambda\sigma}$ in the presence of magnetic charges was first
considered$^{23}$ relevant to chiral symmetry.
The effect of a similar U(1) $\theta$ term was discussed for the purpose of
considering the induced electric charges$^{24}$ as quantum excitations of
dyons associated with the 't Hooft Polyakov monopole and generalized magnetic
monopoles$^{20,24}$. For our purpose, We expect that if a QED $\theta$
term is included, it may possiblely be an indication of the unification.
A $\theta$ term needs to be included in the unification gauge theory for
$\Pi_3(G)=Z$ for the unification group G, magnetic monopoles with charges
involving the QED U(1) symmetry are generated through the spontaneous gauge
symmetry breaking. Generally, such an induced $\theta$ term in QED may not
be discarded in the presence of magnetic monopoles.

As a remark, our quantization rule can also be obtained by using constraints of
Gauss' law. This will be given elsewhere.

I thank Y. S. Wu and A. Zee for valuable discussions.


\newpage
\begin{thebibliography}{99}
\bibitem{thfcnc}C. N. Yang and R. L. Mills, Phys. Rev. 96, 191 (1954).
\bibitem{thfcnc}A. A. Belavin, A. M. Polyakov, A. S. Schwarz and Yu. Tyupkin,
Phys. Lett. 59B, 85 (1975); R. Jackiw and C. Rebbi, Phys. Rev. Lett. 37, 172
(1976); C. Callan, R. Dashen and D. Gross, Phys. Lett. 63B, 334 (1976);
G. 't Hooft, Phys. Rev. Lett. 37, 8 (1976) and Phys. Rev. D14, 334(1976).
\bibitem{thfcnc}R. Peccei, in "CP Violations" (ed. C. Jarlskog, World
Scientific
, Singapore, 1989).
\bibitem{thfcnc}R. D. Peccei and H. R. Quinn, Phys. Rev. Lett. 38, 1440 (1977);
Phys. Rev. D16, 1791 (1977).
\bibitem{thfcnc}S. Weinberg, Phys. Rev. Lett. 40, 223 (1978); F. Wilczek, Phys.
Rev. Lett. 40, 271 (1978).
\bibitem{thfcnc}Y. S. Wu and A. Zee, Nucl. Phys. B258, 157 (1985).
\bibitem{thfcnc}R. Jackiw, in "Relativity, Groups and Topology II", (Les
Houches
1983, ed. B. S. DeWitt and R. Stora, Noth-Holland, Amsterdam, 1984).
\bibitem{thfcnc}P. A. M. Dirac, Proc. Roy. Soc. A133, 60 (1931); Phys. Rev. 74,
817 (1948); 't Hooft, Nucl. Phys. 79, 276 (1974); A. M. Polyakov, JETP Lett.
20,
 194 (1974). For a review, see S. Coleman, in "The Unity of the Fundamental
Interactions" and references theirin.
\bibitem{thfcnc}T. T. Wu and C. N. Yang, Phys. Rev. D12, 3845 (1975).
\bibitem{thfcnc}For a brief review, see for example, B. Zumino, in "Relativity,
Groups and Topology II", (Les Houches 1983, ed. B. S. DeWitt and R. Stora,
Noth-Holland, Amsterdam, 1984)
\bibitem{thfcnc}See for example, S. T. Hu, "Homotopy Theory", (Academic Press,
New York, 1956); N. Steenrod, "The topology of Fiber Bundles", (Princeton Univ.
Press, Princeton, NJ 1951).
\bibitem{thfcnc}E. Witten, Phys. Lett. B117, 324 (1982).
\bibitem{thfcnc}S. Elitzur and V. P. Nair, Nucl. Phys. B243, 205 (1984).
\bibitem{thfcnc}S. Okubo, H. Zhang, Y. Tosa, and R. E. Marshak, Phys. Rev. D37,
 1655 (1988).
\bibitem{thfcnc}H. Zhang, S. Okubo, and Y. Tosa, Phys. Rev. D37, 2946 (1988).
\bibitem{thfcnc}H. Zhang and S. Okubo, Phys. Rev. D38, 1800 (1988).
\bibitem{thfcnc}S. Okubo and H. Zhang, in "Perspectives on Particle Physics
(ed. S. Matsuda, T. Muta, and R. Sakaki, World Scientific, Singapore, 1989).
\bibitem{thfcnc}A. T. Lundell and Y. Tosa, J. Math. Phys. 29, 1795 (1988).
\bibitem{thfcnc}S. Okubo and Y. Tosa, Phys. Rev. D40, 1925 (1989).
\bibitem{thfcnc}See P. Goddard and D. Olive, Rep. Prog. Phys. 41, 1375 (1978);
P. Goddard, J. Nuyts, D. Olive, Nucl. Phys. B125, 1 (1977).
\bibitem{thfcnc}C. Dokos and T. Tomaras, Phys. Rev. D21, 2940 (1980).
\bibitem{thfcnc}J. Preskill, Phys. Rev. Lett. 43 1365 (1979);
G. Giacomelli, in "Monopoles in Quantum Field Theory", (World Scientific,
Singapore, 1981).
\bibitem{thfcnc}H. Pagels, Phys. Rev. D13, 343(1976); W. Marciano and H.
Pagels,
Phys. Rev. D14, 531(1976).
\bibitem{thfcnc}E. Witten, Phys. Lett. B86, 283 (1979); E. Tomboulis and G. Woo
, Nucl. Phys. B107, 221 (1976); H. Zhang, Phys. Rev. D36, 1868 (1987).
\end{thebibliography}
\end{document}